\documentclass[letter,superscriptaddress,twocolumn,prl,showkeys,showpacs]{revtex4-1}

\usepackage[utf8]{inputenc}
\usepackage{amsmath}   
\usepackage{graphicx}   
\usepackage{verbatim}  
\usepackage{color}     
\usepackage{subfigure}  
\usepackage{hyperref}   
\usepackage{natbib}
\usepackage{gensymb}
\usepackage{color}
\usepackage{epstopdf}
\usepackage{multirow}
\usepackage{xfrac}
\begin{document}

\title{Local Real-Space View of the Achiral 1$T$-TiSe$_2$ 2 $\times$ 2 $\times$ 2 Charge Density Wave}

\author{B. Hildebrand}
\altaffiliation{Corresponding author.\\ baptiste.hildebrand@unifr.ch}
\affiliation{D{\'e}partement de Physique and Fribourg Center for Nanomaterials, Universit{\'e} de Fribourg, CH-1700 Fribourg, Switzerland}

\author{T. Jaouen}
\altaffiliation{Corresponding author.\\ thomas.jaouen@unifr.ch}
\affiliation{D{\'e}partement de Physique and Fribourg Center for Nanomaterials, Universit{\'e} de Fribourg, CH-1700 Fribourg, Switzerland}

%\author{C. Didiot}
%\affiliation{D{\'e}partement de Physique and Fribourg Center for Nanomaterials, Universit{\'e} de Fribourg, CH-1700 Fribourg, Switzerland}

%\author{E. Razzoli}
%\affiliation{D{\'e}partement de Physique and Fribourg Center for Nanomaterials, Universit{\'e} de Fribourg, CH-1700 Fribourg, Switzerland}
\author{M.-L. Mottas}
\affiliation{D{\'e}partement de Physique and Fribourg Center for Nanomaterials, Universit{\'e} de Fribourg, CH-1700 Fribourg, Switzerland}

\author{G. Monney}
\affiliation{D{\'e}partement de Physique and Fribourg Center for Nanomaterials, Universit{\'e} de Fribourg, CH-1700 Fribourg, Switzerland}

%\author{F. Vanini}
%\affiliation{D{\'e}partement de Physique and Fribourg Center for Nanomaterials, Universit{\'e} de Fribourg, CH-1700 Fribourg, Switzerland}

\author{C. Barreteau}
\affiliation{Department of Quantum Matter Physics, University of Geneva, 24 Quai Ernest-Ansermet, 1211 Geneva 4, Switzerland}

%\author{A. Ubaldini}
%\affiliation{Department of Quantum Matter Physics, University of Geneva, 24 Quai Ernest-Ansermet, 1211 Geneva 4, Switzerland}

\author{E. Giannini}
\affiliation{Department of Quantum Matter Physics, University of Geneva, 24 Quai Ernest-Ansermet, 1211 Geneva 4, Switzerland}

%\author{H. Berger}
%\affiliation{Institut de G{\'e}nie Atomique, Ecole Polytechnique F{\'e}d{\'e}rale de Lausanne, CH-1015 Lausanne, Switzerland}

\author{D. R. Bowler}
\affiliation{London Centre for Nanotechnology and Department of Physics and Astronomy, University College London, London WC1E 6BT, UK}

\author{P. Aebi}
\affiliation{D{\'e}partement de Physique and Fribourg Center for Nanomaterials, Universit{\'e} de Fribourg, CH-1700 Fribourg, Switzerland}

\begin{abstract}

The transition metal dichalcogenide 1$T$-TiSe$_2$ is a quasi-two-dimensional layered material undergoing a commensurate 2 $\times$ 2 $\times$ 2 charge density wave (CDW) transition with a weak periodic lattice distortion (PLD) below $\approx$ 200 K. Scanning tunneling microscopy (STM) combined with intentionally introduced interstitial Ti atoms allows to go beyond the usual spatial resolution of STM and to intimately probe the three-dimensional character of the PLD. Furthermore, the inversion-symmetric, achiral nature of the CDW in the $z$-direction is revealed, contradicting the claimed existence of helical CDW stacking and associated chiral order. This study paves the way to a simultaneous real-space probing of both charge and structural reconstructions in CDW compounds.

\end{abstract}
\date{\today}
\maketitle

Transition metal dichalcogenides (TMDCs) have been extensively studied for decades, but recently, the development of methods to obtain nanosheets or even monolayers has brought renewed interest for these materials because of their attractive electronic and optoelectronic properties for applications \cite{Chhowalla2013a, Wang2012}. Many TMDCs also undergo charge density waves (CDW) or/and superconducting phase transitions at low temperature \cite{Withers1986, CastroNeto2001, Fang2005a, Chatterjee2015, Joe2014} and recently, 1$T$-TiSe$_2$ has been even proposed as the first TMDC exhibiting a CDW with a scalar chiral order \cite{Ishioka2010a, Castellan2013}. 

Scanning tunneling microscopy (STM) probes the real-space surface electron density. Therefore, the electronic reconstruction associated with the commensurate 2 $\times$ 2 $\times$ 2 CDW of  1$T$-TiSe$_2$ can be easily tracked in the topmost Se layer up to atomic resolution \cite{Slough1988,novello2015}. This local approach has indeed allowed to propose the chirality of the CDW \cite{Ishioka2010a, Iavarone2012a} and to observe the formation of phase-shifted CDW domains induced by intercalation of Ti or Cu atoms\cite{Hildebrand2016,Yan2017,Hildebrand2017,Novello2017}. However, the three-dimensional (3D) character of the CDW as well as the periodic lattice distortion (PLD) accompanying the phase transition can, in principle, not be directly probed, the displacement amplitude induced by the PLD \cite{salvo1976,Fang2017} being well below the usual spatial resolution of STM \cite{chen2008}. 

Yet, being able to obtain information about the 3D character of the CDW as well as getting insight about the PLD in real-space is of fundamental interest in the context of domain formation or CDW dimensional crossover \cite{Chen2016}, but also with respect to the question of chirality which strictly signifies a loss of inversion symmetry between two adjacent Se-Ti-Se sandwiches. 

Here, the electronic signature of defects induced by the presence of interstitial Ti atoms in-between Se-Ti-Se sandwiches is exploited for probing the symmetry of the PLD. This combined STM-defect technique therefore goes beyond the resolution limitation of STM providing access to the structural and intimate 3D nature of the 1$T$-TiSe$_2$ CDW. Interstitial Ti defects are first recognized and characterized. A systematic asymmetry induced by the surrounding surface PLD is observed and the associated deformation is uniquely determined with the help of density functional theory (DFT) calculations. Furthermore, the potential of this new technique is highlighted through a simultaneous measurement of two adjacent Se-Ti-Se sandwiches at a step edge where the CDW/PLD is found to display the expected commensurate 2 $\times$ 2 $\times$ 2 signature with atomic displacements as proposed by Di Salvo \textit{et al.} \cite{salvo1976}. Therefore, there is no evidence of a loss of inversion symmetry, contradicting recent claims of a chiral phase \cite{Ishioka2010a, Iavarone2012a}.

The strength of this combined STM-defect technique also relies on its relative simplicity. Indeed, a defect density of less than 1$\%$ allows precise identification of the surface PLD. Hence, this work not only paves the way towards a better local understanding of structural reconstructions in CDW compounds but is also highly promising for phase transition studies in general, where a local and real-space vision of entangled electronic and structural instabilities is essential.

The 1$T$-TiSe$_2$ single crystals were grown by iodine vapor transport at 700 \celsius$~$with 5$\%$ additional Ti in the growth tube with respect to perfect stoichiometric conditions. Resistivity measurements were performed by a standard four-probe method using a lock-in as current source and voltage meter. The samples were cleaved in-situ below 10$^{-7}$ mbar at room temperature. Constant current STM images were recorded at 4.6 K using an Omicron LT-STM, with bias voltage V$_{\text{bias}}$ applied to the sample. Base pressure was better than 5$\times$10$^{-11}$ mbar.

DFT calculations have been performed using the plane-wave pseudopotential code VASP \cite{Kresse1993, Kresse1996}, version 5.3.3. Projector augmented waves \cite{Kresse1999} were used with the Perdew-Burke-Ernzerhof (PBE) \cite{Perdew1996} exchange correlation functional. The cell size of our model was 28.035 \AA~$\times$ 28.035 \AA. The 1$T$-TiSe$_2$ surface was modeled with two layers and the bottom Se layer fixed. A Monkhorst-Pack mesh with 2 $\times$ 2 $\times$ 1 $k$ points was used to sample the Brillouin zone of the cell. The parameters gave an energy difference convergence of better than 0.01 eV. During structural relaxations, a tolerance of 0.03 eV/\AA~ was applied. STM images were generated using the Tersoff-Hamann approach \cite{Tersoff1983} in which the current $I(V)$ measured in STM is proportional to the integrated LDOS of the surface using the bSKAN code \cite{Hofer2003}.

\begin{figure}[t]
\includegraphics[scale=1]{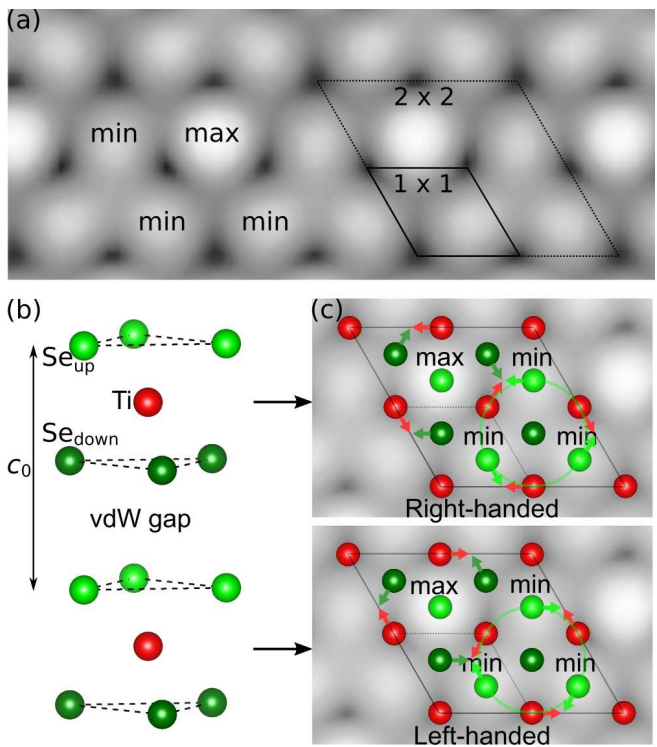}
\caption{(a) Simulation of the CDW as observed in STM at  $V_{\text{bias}}$= 100 mV with indication of three min. and one max. of the charge modulation composing the 2 $\times$ 2 surface unit cell. The black and the dashed rhombi indicate the 1 $\times$ 1 and 2 $\times$ 2 unit cells, respectively. (b) Side view of two 1$T$-TiSe$_2$ layers separated by a van der Waals (vdW) gap. The out-of-plane $c_0$ lattice constant of the 1 $\times$ 1 $\times$ 1 unit cell is also shown. (c) Schematic top view of the 2 $\times$ 2 $\times$ 2 PLD corresponding to the two adjacent layers shown in (b). The arrows indicate the directions of displacement of the atoms after transition to the CDW state. The underlaying modulation shows the registry of the min. and max. of the Se$_{\text{up}}$ layer with respect to the PLD. The two possible orientations of displacement are denominated right-handed and left-handed, respectively. }\label{fig1}
\end{figure}

Figure \ref{fig1} (a) shows a DFT simulation, relaxed from displacements following Di Salvo \textit{et al.} \cite{salvo1976}, of a high-resolution low-temperature empty-state STM image of 1$T$-TiSe$_2$ close to $E_F$ (small positive $V_{\text{bias}}$) \cite{novello2015}. In addition to the 1 $\times$ 1 charge density corresponding to the position of the atoms in the topmost Se layer \cite{Slough1988}, the 2 $\times$ 2 CDW modulation can be identified through the succession of atoms appearing more or less intense and labeled CDW max. [see max in Fig. \ref{fig1} (a)] and CDW min [see min in Fig. \ref{fig1} (a)]. The 2 $\times$ 2 surface unit cell contains exactly one CDW max. and three CDW min. [dashed rhombus in Fig. \ref{fig1} (a)]. In-plane displacement amplitude of the Se atoms [see Se$_{\text{up}}$ and Se$_{\text{down}}$ in Fig. \ref{fig1} (b)] corresponds to 0.028 \AA$~$ at 77 K \cite{salvo1976} which is almost two orders of magnitude below the highest lateral resolution reached by standard STM and can thus not be directly tracked with this technique. Interestingly, according to the PLD, the new unit cell also contains exactly one Se$_{\text{up}}$ atom which has not moved with respect to its normal state position and three Se$_{\text{up}}$ which have undergone a small distortion. This observation demonstrates the close relationship between CDW modulation and PLD and allows to directly associate the CDW max. to the non-displaced Se atoms [see Se$_{\text{up}}$ max. atoms in Fig. \ref{fig1} (c)] and the CDW min. to the displaced ones [see arrows on Se$_{\text{up}}$ min. atoms in Fig. \ref{fig1} (c)]. 

Nevertheless, we would like to stress that, measuring the 2 $\times$ 2 charge modulation with STM does not allow to uniquely determine the \textit{handedness} of the PLD occuring within the first Se-Ti-Se sandwich. Indeed, due to the characteristic antiphase locking of the 2 $\times$ 2 $\times$ 2 CDW between adjacent Se-Ti-Se layers [see Fig. \ref{fig1} (b) and (c)] \cite{salvo1976}, there are two possible periodic distortions [\textit{right-handed} or \textit{left-handed}, see Fig. \ref{fig1} (c)] for one CDW modulation at the surface. STM alone is therefore completely blind to this essential parameter exhibiting the 3D character of the CDW.

\begin{figure*}[t]
\includegraphics[scale=1]{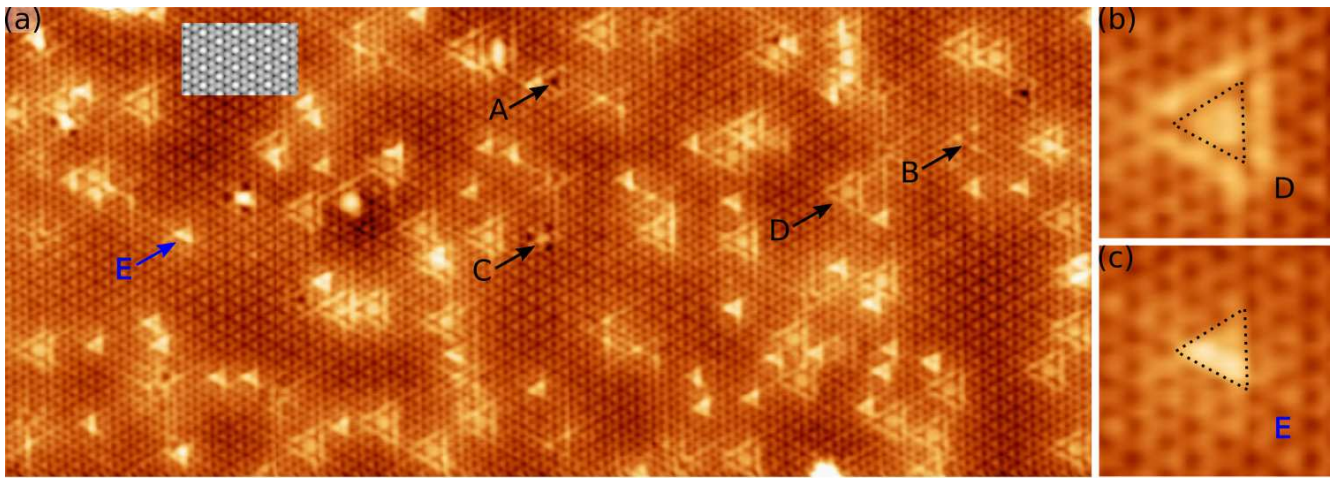}
\caption{(a) 40 $\times$ 17 nm$^2$ constant current STM image of the 700\celsius-grown 1$T$-TiSe$_2$ sample with 5$\%$ additional Ti in the growth tube. $V_{\text{bias}}$= 150 mV, $I$= 0.2 nA. Native defects are labeled A-D according to the reference \cite{Hildebrand2014}. The new defect is labeled E. An inset showing the DFT-simulated STM image from Fig. \ref{fig1} (a) is added as an eye guide; note that the image has been rotated by 30° with respect to Fig. \ref{fig1} (a). (b), (c) zooms-in on one defect of type D, and E, respectively. Dotted triangles are added for highlighting the orientation of three Se atoms concerned with the electronic signatures of the defects.}\label{fig2}
\end{figure*}

Figure \ref{fig2} (a) displays a measured large empty-state STM image close to $E_F$ (150 mV). Many types of atomic defects can be recognized, the overall impurity density approximately corresponding to $\approx$2$\%$. A large proportion of defects are the well-known native impurities described in a previous study \cite{Hildebrand2014}. Defect A corresponds to a Se vacancy, defects B and C are Se$_{\text{down}}$ substitutions by iodine and oxygen atoms, respectively, and defect D is associated to intercalated Ti in the vdW gap. Based on surveys from STM images obtained at different regions of the sample (not shown), the density of intercalated Ti defects which is mostly determined by the growth temperature (700 \celsius) \cite{salvo1976}, is 0.57$\pm$0.07$\%$ and has therefore no impact on the CDW long-range phase coherence as seen Fig. \ref{fig2} (a) \cite{Hildebrand2016}. 

In addition to these four native impurities, the non-stoichiometric growth conditions have introduced another defect, never observed previously \cite{Hildebrand2014, novello2015}. It stands out from the periodic and ordered Se layer as small, bright asymmetric triangle [defect E, Fig. \ref{fig2} (a)]. The estimated density of defect E is 0.79$\pm$0.08$\%$, a value close to the density of intercalated Ti [defect D, Fig. \ref{fig2} (a)]. It should not be confused with the latter, which also presents a very similar small triangular shape due to enhanced density of states on top of the three neighboring Se$_{\text{up}}$ atoms in filled-states \cite{Hildebrand2014}, below -0.2 V. One has to keep in mind that at this bias voltage (0.15 V), i.e. in empty-states, the electronic perturbation induced by intercalated Ti is spatially much more extended, in such a way that the difference between defect D and E is manifest [see zooms-in on both defects in Fig. \ref{fig2} (b) and (c) for comparison]. Figure \ref{fig2} (c) also allows us to highlight the particular asymmetric nature of the electronic signature of defect E with respect to the lattice. Indeed, two of the three topmost Se atoms concerned with its electronic perturbation are clearly brighter than the third one building a "bright edge".

\begin{figure}[t]
\includegraphics[scale=1]{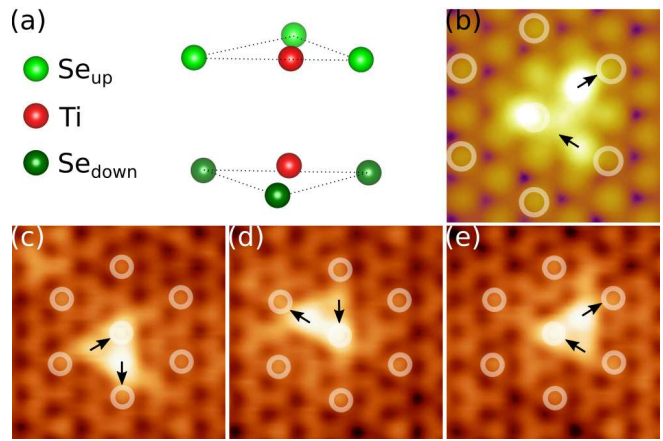}
\caption{(a) Side view of the DFT-calculated relaxed structure with an interstitial Ti atom. (b) DFT simulated STM image at $V_{\text{bias}}$= 150 mV. White circles mark the CDW max. and arrows show the direction of simulated PLD displacements according to Di Salvo \textit{et al.} \cite{salvo1976} of the Se$_{\text{up}}$ atoms concerned with the electronic perturbation of the defect. (c)-(e) Zoom-ins on three defects of type E from Fig. \ref{fig2} with three different orientations. White circles again mark the CDW max. extended from the defect-free region and arrows show the direction of PLD displacements deduced from (b).}\label{fig3}
\end{figure}

Given the special growth conditions, defect E can be reasonably attributed to additional Ti atoms. Also, the orientation of the triangular shape of defect E is identical to the central triangle of the intercalated-Ti electronic signature [see small dotted triangles Fig. \ref{fig2} (b) and (c)]. This indicates that, as intercalated Ti, defect E is in vertical alignment with a structural Ti atom. Therefore, we performed DFT simulations of STM images in the 2 $\times$ 2 $\times$ 2 phase with an interstitial Ti atom located inside a TiSe$_6$ octahedron forming a Ti$_2$Se$_6$ structure [Fig. \ref{fig3} (a))]. As seen in Fig. \ref{fig3} (b), the DFT-simulated image for the relaxed structure at 150 mV of $V_{\text{bias}}$ is in excellent agreement with our measurement and especially accounts for the characteristic bright edge. The arrows give the direction of displacements of the top Se atoms according to the PLD introduced in the simulation.      
 
In particular, one of the atoms building the bright edge is a CDW max. (non-displaced atom) and the second bright atom corresponds to an atom which moves away from the first one according to the PLD. Comparing the simulation to the three observed conformations of the bright edge induced by interstitial Ti atoms in Fig. \ref{fig2} [see zoom-ins in Fig. \ref{fig3} (c)-(e)] first confirms that it is always constituted of one CDW max. \footnote{Interstitial Ti atoms can also be situated in such a way that the three atoms composing its triangular shape are three min. of the CDW modulation. Even though, a detailed analysis of this conformation has been done, it is not presented here since the measurement as well as the calculation show that, for this specific case, the asymmetry is only weak and not well defined due to the perfectly symmetric nature of its position with respect to the CDW/PLD. Hence, in this position with respect to the lattice, interstial Ti can not be used as a probe of the PLD. For more precisions about the PLD-induced inequivalent conformations of single-atom defects see Refs.\cite{novello2015, Hildebrand2017}.}. Then, plotting the corresponding directions of distortion on the observed conformations [arrows Fig. \ref{fig3} (c)-(e)] uniquely demonstrates that, here, the underlying PLD is the left-handed one [Fig. \ref{fig1} (c)]. 

\begin{figure}[t]
\includegraphics[scale=1]{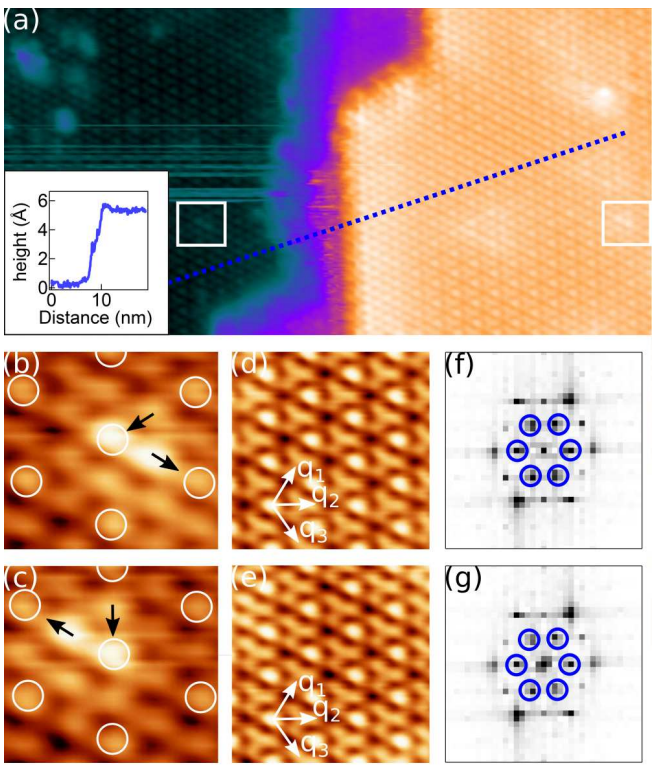}
\caption{(a) 20 $\times$ 10 nm$^2$ constant current STM image at $V_{\text{bias}}$= 150 mV and $I$= 0.1 nA of two adjacent $1T$-TiSe$_2$ layers. Inset: height difference of approximately 6 \AA$~$ corresponding to the separation between two Se-Ti-Se sandwiches obtained along the profile represented by the blue dotted line. Two white rectangles indicate the presence of two defects induced by interstitial Ti, one on each layer. (b)-(c) zooms-in on the defects of the bottom and top layer respectively with white circles indicating the position of the maxima of the CDW charge modulation and black arrows indicating the corresponding PLD-induced atomic displacements. (d)-(e) 2.8  $\times$ 2.8 nm$^2$ zooms-in showing the CDW modulation in the bottom and top layer respectively with arrows indicating the orientations of the CDW $q$-vectors. (f)-(g) FFT-amplitude plots obtained from (d) and (e) with blue circles highlighting the extra spots originating from the CDW charge modulation.}\label{fig4}
\end{figure}

One has to realize that one single interstitial-Ti atom defect on a STM image is already sufficient for probing the orientation of the underlying PLD ensuring, in turn, that the 3D character of the CDW can be probed through this novel method with minimal defect-induced disturbance. Figure \ref{fig4} (a) shows an STM image at a step edge between two adjacent layers separated by $\approx 6$ \AA$~$corresponding to one $c_0$ lattice constant [see inset of Fig. \ref{fig4} (a)]. The CDW charge modulation is recognizable on both terraces and can be easily tracked. Also, two interstitial-Ti defects are clearly observable on each side of the step edge [see white rectangles in Fig. \ref{fig4} (a)] therefore providing atomic probes of the PLD orientation for the top and bottom TiSe$_2$ layers. Comparing the positions of the CDW maxima with the electronic signatures of the interstitial-Ti atoms for both layers [see white circles on zooms-in Fig. \ref{fig4} (b) and (c)] allows for determining the direction of motion that the Se atoms have undergone through the phase transition [see black arrows in Fig. \ref{fig4} (b) and (c)]. The PLD is therefore found to be left-handed for the top layer and right-handed for the bottom [Fig. \ref{fig1} (c)], demonstrating that we are facing a local real-space view of the 2 $\times$ 2 $\times$ 2 character of the 1$T$-TiSe$_2$ CDW.

Furthermore, our experiment allows to readily conclude on the inversion-symmetric achiral nature of the CDW as initially proposed by Di Salvo \textit{et al.} \cite{salvo1976}. Indeed, the recently claimed chirality of the 1$T$-TiSe$_2$ CDW is based on a helical stacking of the three CDW $q$-vectors along the $z$-direction with a 2$c$/3 interval, leading to the existence of so-called virtual layers with shifted CDW density peaks \cite{Ishioka2010a}. This implies different amplitudes of CDW charge modulations along the three $q$-vectors not only in a single layer but also between two $c$-separated TiSe$_2$ layers.   
Two identically sized defect-free regions of both layers are selected in Fig. \ref{fig4} (a) and shown in Fig. \ref{fig4} (d) and (e). Their close similarity in real-space already suggest that no relative phase shifts exist between the three CDW $q$-vectors. In addition, the integrated intensities of the CDW extra-spots on fast-Fourier transformed (FFT) images obtained from Fig. \ref{fig4} (d) and (e) [Fig. \ref{fig4} (f) and (g)] show a negligible variation of less than 5$\%$. This definitely confirms the conventional layer stacking in 1$T$-TiSe$_2$ and rules out the proposed 2$c$/3 helical CDW stacking of chiral CDW phases \cite{Ishioka2010a}.

In summary, the combination of STM with a low-density of specific buried defects has been used to uncover the periodic lattice distortion at the 1$T$-TiSe$_2$ surface. The potential of this new PLD-sensitive probe has been then exemplified through the measurement of the inversion of its handedness at a step edge therefore providing a real-space view of the 2 $\times$ 2 $\times$ 2 character of the CDW. The inversion-symmetric nature of the CDW in the $z$-direction has been finally confirmed, contradicting the existence of CDW helical stacking and associated chiral order. In principle, this new method could be easily extended to the analysis of the mixed CDW-superconducting state in Cu-intercalated 1$T$-TiSe$_2$ \cite{Yan2017}, as well as to other materials exhibiting phase transitions for which a better understanding of the interplay between electronic and structural degrees of freedom is required.

\begin{acknowledgments}
This project was supported by the Fonds National Suisse pour la Recherche Scientifique through Div. II. We would like to thank C. Monney, C. Renner and M. Spera for motivating discussions. Skillful technical assistance was provided by F. Bourqui, B. Hediger and O. Raetzo.

B.H. and T.J. equally contributed to this work.
\end{acknowledgments}

\bibliography{library1}
\end{document}